\documentclass{osa-article}

\journal{oe}

\articletype{Research Article}

\usepackage{lineno}
\usepackage{siunitx}

\begin{document}

\title{The effect of rotational Raman response on ultra-flat supercontinuum generation in gas-filled hollow-core photonic crystal fibers}

\author{Mohammed Sabbah,\authormark{1,*} Federico Belli,\authormark{1} Christian Brahms,\authormark{1}  and John C. Travers, \authormark{1}}

\address{\authormark{1}{School of Engineering and Physical Sciences, Heriot-Watt University, Edinburgh, UK}}

\email{\authormark{*}M.Sabbah@hw.ac.uk} 



\begin{abstract}
We experimentally and numerically investigate flat supercontinuum generation in gas-filled anti-resonant guiding hollow-core photonic crystal fiber. By comparing results obtained with either argon or nitrogen we determine the role of the rotational Raman response on the supercontinuum formation. When using argon, a supercontinuum extending from \SI{350}{\nm} to \SI{2}{\micro\meter} is generated through modulational instability. Although argon and nitrogen exhibit similar Kerr nonlinearity and dispersion, we find that the energy density of the continuum in the normal dispersion region is significantly lower when using nitrogen. Using numerical simulations, we find that due to the closely spaced rotational lines in nitrogen, gain suppression in the fundamental mode causes part of the pump pulse to be coupled into higher-order modes which reduces the energy transfer to wavelengths shorter than the pump.
\end{abstract}

\section{Introduction}
Modulational instability (MI) in optical fibers enables the use of continuous-wave or long-pulse pump lasers to drive ultrafast soliton dynamics, leading to extreme spectral broadening and the formation of a supercontinuum \cite{Hasegawa1986,dudley_supercontinuum_2006,genty_fiber_2007}. This has been widely explored in solid-core fibers and is the basis for commercial supercontinuum sources, which offer a wide range of applications in biomedical imaging, spectroscopy, and chemical sensing \cite{alfano_supercontinuum_2016,dudley_taylor_2010}. Over the last decade, it has been shown that gas-filled hollow-core fibers provide a rich system for exploring nonlinear dynamics. In particular, they significantly expand the guidance window to the ultraviolet and mid-infrared spectral region, and enable the tuning of both the dispersion and nonlinearity through the gas pressure \cite{travers_ultrafast_2011, russell_hollow-core_2014}. Furthermore, even the fundamental character of the nonlinear response can be altered through the type of gas used and the role of plasma or Raman effects \cite{markos_hybrid_2017}.

In MI-driven supercontinuum formation, a high-power pump causes an exponential amplification of sidebands symmetrically located around the pump frequency and lying beyond the pump bandwidth. For high peak power and weak dispersion, these sidebands can already be very widely spaced. In the time domain this sideband formation corresponds to the breakup of the pump field into ultrafast optical solitons and remnant non-solitonic radiation. Subsequent collisions and interactions between all of these fields, mediated by cross-phase modulation (XPM), provide a mechanism for enhancing the spectral bandwidth and the formation of broader and flatter supercontinua~\cite{Genty:04, skryabin_theory_2005}.

The potential of using gas-filled hollow-core fibers for MI-driven supercontinuum was first outlined by Travers \textit{et al.}~in 2011 \cite{travers_ultrafast_2011}, and experimentally demonstrated by Tani \textit{et al.}~in xenon-filled Kagomé-style antiresonant hollow-core fibers~\cite{tani_phz-wide_2013}. They achieved a continuum spanning from \SI{320}{\nm} to~\SI{1300}{\nm}. In noble-gas filled fibers there is no Raman response and so the broadening and flattening of the continuum in the work of Tani \textit{et al.}~was entirely mediated by XPM. Mousavi \textit{et al.}~used a fiber with a similar design, but filled with ambient air, to study the additional effect of the Raman response on the propagation of picosecond pulses, and observed the formation of a continuum ranging from \SI{900}{\nm} up to \SI{1500}{\nm} \cite{mousavi_nonlinear_2018}.  Recently, a new supercontinuum mechanism was proposed and demonstrated, based on the broadening and merging of the vibrational Raman frequency comb \cite{gaoRamanFrequencyCombs2022}. Using that technique, a supercontinuum spanning from \SI{440}{\nm} to~\SI{1200}{\nm} in a nitrogen-filled fiber was achieved. A similar approach using liquid-filled capillaries has also been demonstrated~\cite{Fanjoux:17}, and further extension into the infrared using a deuterium filled fiber has been achieved~\cite{Gladyshev2022}. 

In this paper, we experimentally and numerically study the generation of supercontinua through MI in an anti-resonant hollow-core fiber filled with argon or nitrogen. We generate a broadband, ultra-flat supercontinuum extending from \SI{350}{\nm} up to \SI{2000}{\nm} in both gases, but with significant differences in the energy density in the normal dispersion region. Through detailed numerical simulations we investigate the effect of the Raman response on the MI dynamics and the formation of the supercontinuum in nitrogen. We find that gain suppression due to the closely spaced rotational lines in nitrogen causes energy to be coupled into higher-order modes of the fiber, which reduces the energy transfer to the wavelength region shorter than the pump.  Our results here contribute to the recent and growing interest into the subtle nonlinear dynamics of pulse propagation in hollow waveguides filed with Raman active gas~\cite{tani_generation_2015, belli_vacuum-ultraviolet_2015, safaei_high-energy_2020,Beetar2020}.

\section{Methods}
\subsection{Experimental setup}

\begin{figure}[b!]
\centering
\includegraphics{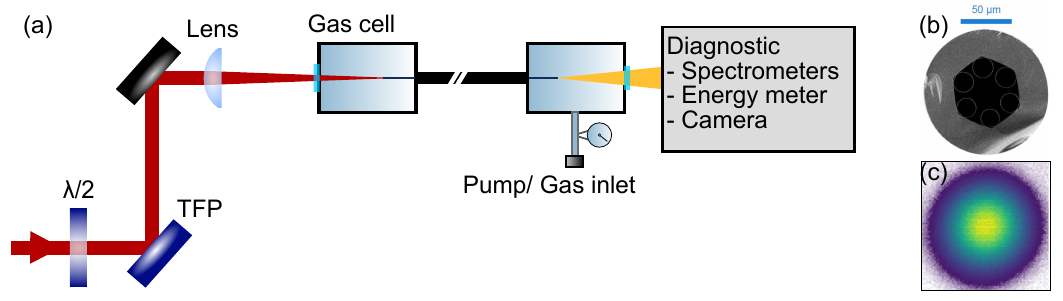}
\caption{The experimental setup. (a) Optical layout of the experiment. TFP: thin-film polariser, $\lambda$/2: half wave-plate. (b) Cross section of the PCF used in the experiment. (c) Near-field beam profile of the output mode of the PCF while in vacuum.}
\label{fig:Setup}
\end{figure}

Figure~\ref{fig:Setup}(a) shows the experimental setup. A \SI{1030}{\nm} pump laser (Light Conversion PHAROS) with tuneable pulse duration (from \SI{220}{\fs} full-width half-maximum (FWHM)  up to \SI{10}{\ps}) is used at \SI{1}~kHz repetition rate. The pulse energy is controlled using a half-wave plate and a thin-film polarizer. The beam is then coupled to the fiber using a \SI{5}{\cm} focal length anti-reflection-coated plano-convex lens. The fiber is \SI{0.62}{\meter} long and sealed inside a tube which connects two gas cells with optical access provided by \SI{6.3}{\mm}-thick fused silica windows. The output spectrum is collected using an integrating sphere and two fiber-coupled spectrometers. The first spectrometer (Avantes ULS2048XL) is used for the range \SI{200}-\SI{1100}{\nm} and the second (Avantes NIR256-2.5) for \SI{1100}-\SI{2500}{\nm}. The whole system is calibrated on an absolute scale with NIST traceable lamps. Figure~\ref{fig:Setup}(b) shows a scanning electron micrograph of the fiber. The single-ring nodeless hollow-core anti-resonant fiber has a core diameter of \SI{32}{\micro\meter} and a wall thickness of \SI{230}{\nm}. According to the anti-resonant model, this wall thickness puts the first high-loss resonance band at around \SI{480}{\nm} \cite{litchinitser_antiresonant_2002, yu_negative_2016}. Apart from this band, the fiber exhibits a broadband, low-loss transmission from \SI{250}{\nm} to above \SI{2}{\micro\meter}.

For MI-based supercontinuum generation, pulses with the same peak power and longer duration produce a smoother and broader spectrum. This is because longer pump pulses contain a larger number of sub-pulses (or modulation periods) produced during MI, all of which contribute to the spectral broadening, resulting in a smoother average supercontinuum. In our experiments we use a pulse duration of \SI{1}{\ps} and a maximum energy of \SI{18}{\micro\joule}, which is more than sufficient to obtain a broad and smooth supercontinuum in our argon experiments.

One advantage of using hollow-core fibers is the ability to change the dispersion profile by simply changing the gas pressure or species, which balances the normal dispersion of the gas with the anomalous dispersion of the waveguide~\cite{travers_ultrafast_2011}. For the generation of a flat supercontinuum across the transmission window of the fiber, the choice of the zero-dispersion wavelength (ZDW) is crucial. For wavelengths shorter than the ZDW, the group velocity is normal, while for longer wavelengths it is anomalous. The ZDW can be shifted to longer wavelengths by increasing the gas pressure. For MI to take place, the pump pulse needs to experience anomalous dispersion. Furthermore, to allow an efficient transfer of energy to the normal dispersion region, the ZDW needs to be close to the pump wavelength~\cite{dudley_supercontinuum_2006, travers_visible_2008}.
When the pulse breaks up into sub-pulses due to MI, the sub-pulses eventually form solitons and dispersive waves (remnant non-solitonic radiation). Some of the dispersive waves formed on the normal-dispersion side travel with similar group velocity to the solitons on the anomalous-dispersion side. These dispersive waves can become trapped by the refractive-index well that is formed by the soliton \cite{genty_route_2005, beaud_ultrashort_1987}. In the absence of Raman nonlinearity, the solitons interact with the dispersive waves through XPM, causing the dispersive waves to shift towards shorter wavelengths. To conserve the photon number during this process, the solitons experience a redshift \cite{tani_phz-wide_2013}. In this way, soliton trapping and XPM lead to the generation of broader supercontinua. The supercontinuum expansion due to this process is eventually restricted by the limits of group-velocity matching between the most red-shifted solitons and shortest wavelength dispersive-waves. This is usually optimised by shifting the ZDW slightly further from the pump. The optimal dispersion for both initiating the supercontinuum, and ensuring the broadest bandwidth for the given fiber transmission window, is obtained when the ZDW is chosen to balance all of these considerations.

\begin{figure}[t!]
\centering
\includegraphics{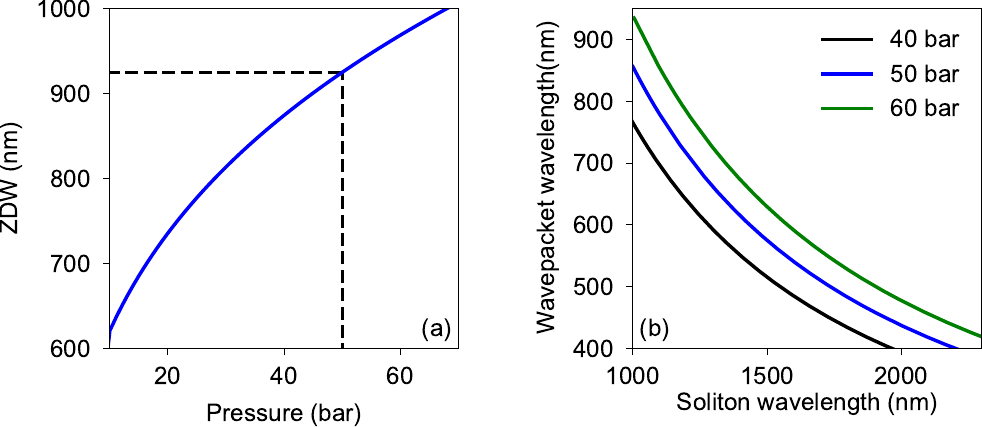}
\caption{(a) The ZDW for increasing pressure using \SI{32}{\micro\meter} core diameter fiber filled with argon. The black dashed lines annotate the experimental parameters. (b) Group velocity matching between IR solitons and dispersive-waves in the normal dispersion for three different pressures. }
\label{fig:GVM_ZDW}
\end{figure}

Figure~\ref{fig:GVM_ZDW}(a) shows how the ZDW changes with argon pressure as calculated using the capillary model~\cite{marcatili_hollow_1964} and the refractive index of argon as given in ref.~\cite{borzsonyi_dispersion_2008}. Figure~\ref{fig:GVM_ZDW}(b) shows the group-velocity matching between the IR soliton wavelength and the dispersive wave wavelength for three gas pressures. In our experiments, we work at \SI{50}{\bar}, which places the ZDW around \SI{895}{\nm}, allowing the first higher-frequency MI sideband to be formed in the normal-dispersion region, around \SI{835}{\nm}, while providing group-velocity matching of the long-wavelength supercontinuum edge around \SI{2}{\um} with the ultraviolet, to at least \SI{450}{\nm} (roughly matching the edges of the fiber transmission window).

\subsection{Numerical model}
Our simulations are based on solving the modal unidirectional pulse propagation equation. The model includes the effect of Kerr, photoionization, and Raman nonlinearities \cite{kolesik_nonlinear_2004, tani_multimode_2014, travers_high-energy_2019, brahms_Luna_2022, ammosov_tunnel_1986, gaoRamanFrequencyCombs2022}. The dispersion is modeled using the Marcatili model \cite{marcatili_hollow_1964}. In our simulations, we  include a range of higher order modes of type HE$_{1m}$, where $m$ is an integer. Quantum noise is added to the input optical field in the simulations through the addition of a noise seed of one photon per mode with a random phase in each spectral discretization bin \cite{drummond_quantum_2001, dudley_supercontinuum_2006, tani_phz-wide_2013}. To match our numerical simulations with experimental results we use \SI{15}{\percent} less pump pulse energy than the corresponding experimental values. This discrepancy could arise due to an imperfect pump pulse shape or an amplified spontaneous emission component from the laser.

\section{Experimental results}
\subsection{Supercontinuum in argon-filled fiber}

\begin{figure}[b!]
\centering
\includegraphics[width=4in]{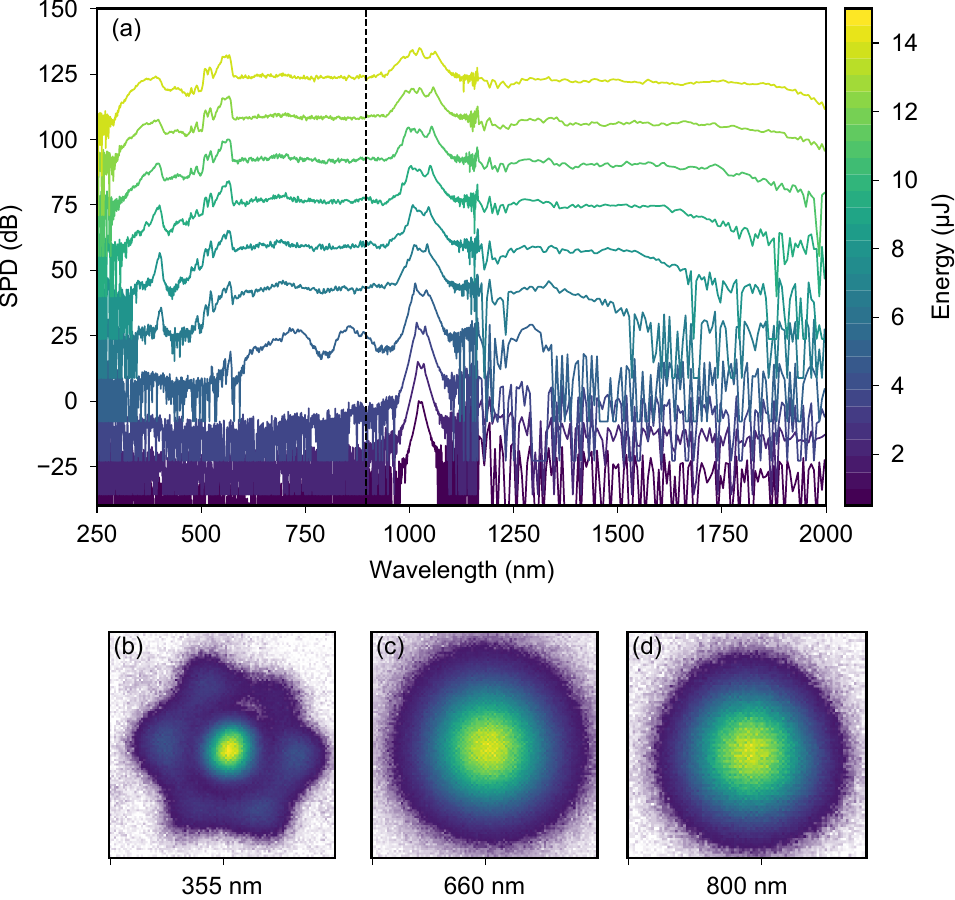}
\caption{(a) Logarithmic spectra for increasing energy with \SI{1.5}{\micro\joule} increments for \SI{50}{\bar} argon and \SI{1}{\ps} pulse duration. The black dashed line indicate the ZDW. (b), (c), and (d) show the measured near-field beam profile at three band-pass filtered wavelengths across the supercontinuum: \SI{355}{\nm}, \SI{660}{\nm}, and \SI{800}{\nm} respectively, for \SI{10}{\micro\joule} pump energy.}
\label{fig:Ar_50bar_escan}
\end{figure}

Figure~\ref{fig:Ar_50bar_escan}(a) shows the experimental evolution of the output spectrum, averaged over 30 pulses, on a logarithmic scale as the input pulse energy is increased in steps of \SI{1.5}{\micro\joule} between \SI{0.5}{\micro\joule} and \SI{14}{\micro\joule}, for \SI{1}{\ps} pulse duration and \SI{50}{\bar} argon pressure. At low energies, MI sidebands appear on both sides of the ZDW, which is labeled by a dashed black line. From around \SI{5}{\micro\joule} input energy, a smooth and broad supercontinuum is formed. As the energy increases further, the continuum becomes exceptionally flat and extends from \SI{350}{\nm} up to \SI{2}{\micro\meter}. The spectral feature around \SI{550}{\nm} is induced by the core-wall resonance, which causes a large spike in the fiber dispersion that enables phase-matched four-wave mixing and resonant dispersive wave emission \cite{tani_effect_2018}. Using a hollow-core fiber with thinner inner capillaries will shift the fiber resonance to shorter wavelengths and allow a smoother continuum in the short-wavelength region. Figure~\ref{fig:Ar_50bar_escan}(b), (c), and (d) show the near-field beam profile across the continuum collected after band-pass filters with \SI{10}{\nm} FWHM bandwidth at \SI{355}{\nm}, \SI{660}{\nm}, and \SI{800}{\nm}, respectively. The mode profiles at \SI{660}{\nm} and \SI{800}{\nm} show Gaussian-like fundamental modes, whereas the UV profile at \SI{355}{\nm} exhibits higher-order mode content. The latter is generated via resonant dispersive-wave emission from the solitons formed during the modulational instability process \cite{tani_phz-wide_2013, tani_multimode_2014}. To verify this and analyse the supercontinuum generation process in more detail, we reproduce these experimental results using numerical simulations. 

\begin{figure}[t!]
\centering
\includegraphics{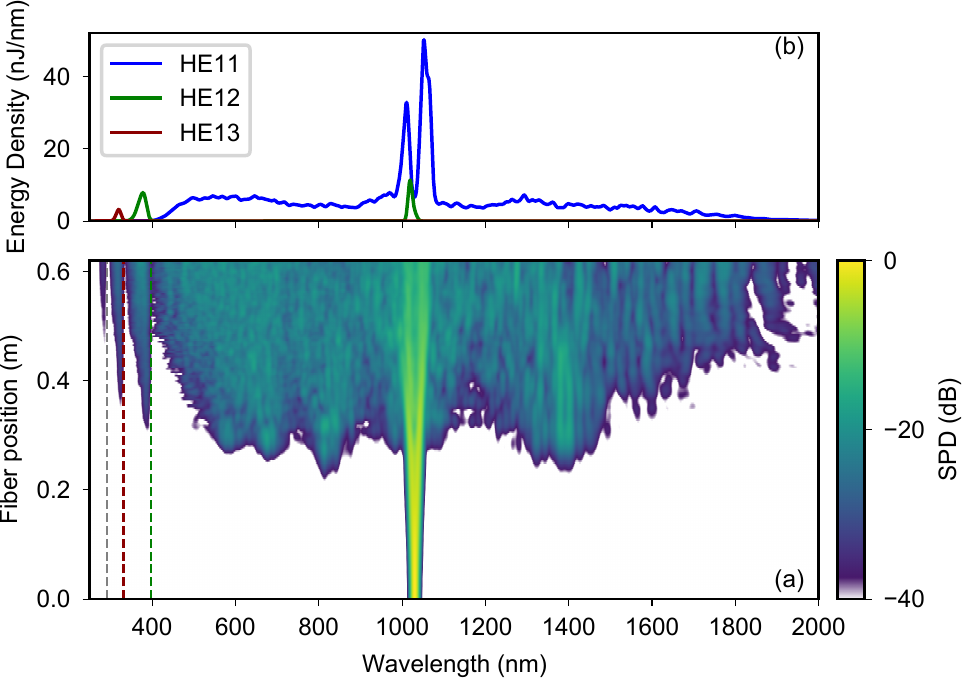}
\caption{Numerical simulation results for a \SI{32}{\micro\meter} core-diameter filled with \SI{50}{\bar} argon, pumped with \SI{1}{\ps} pulses centered at \SI{1030}{\nm} with \SI{8.5}{\micro\joule} input pulse energy. (a) Shows the spectral evolution along the fiber for all modes on a dB scale. (b) The output spectrum at the end of the fiber for the first three HE$_{1m}$ modes processed with a \SI{10}{\nm} Gaussian window and averaged for 30 simulations, each with a different noise seed.}
\label{fig:Ar_sim}
\end{figure}

\begin{figure}[t!]
\centering
\includegraphics{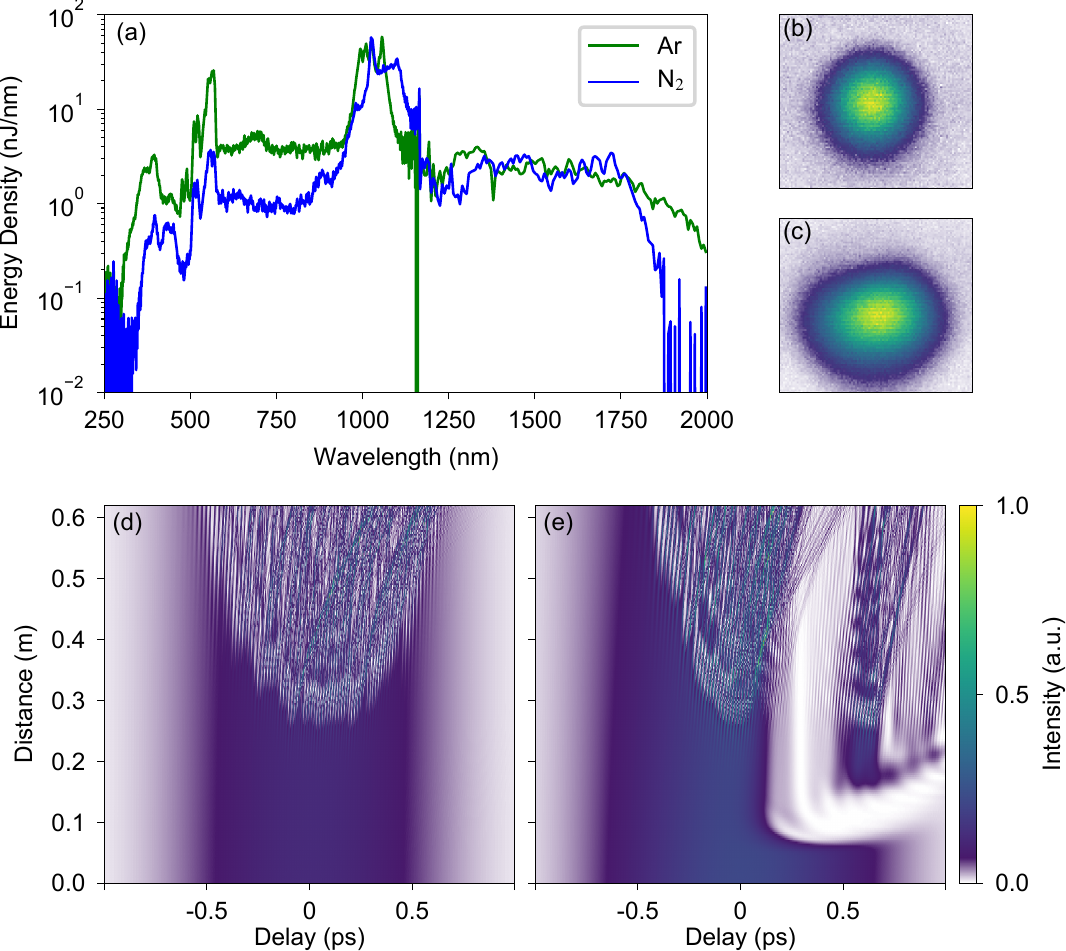}
\caption{(a) Experimental output spectrum for \SI{50}{\bar} argon and nitrogen using \SI{1}{\ps} with \SI{10}{\micro\joule} input pulse energy. (b) Beam profile at \SI{1047}{\nm} for \SI{50}{\bar} argon using \SI{2.2}{\micro\joule}. (c) Beam profile at \SI{1047}{\nm} for \SI{50}{\bar} nitrogen using \SI{2.2}{\micro\joule}. (d) and (e) Show the  simulated temporal evolution of the normalized HE$_{11}$ mode intensity for the same experimental parameters in (a) for both argon and nitrogen respectively.}
\label{fig:ArvsN2}
\end{figure}

Figure~\ref{fig:Ar_sim} shows the numerically simulated output spectrum and spectral evolution along the fiber for the experimental conditions at \SI{8.5}{\uJ} pulse energy. The weak peaks in the UV, which appear around \SI{35}{\cm} into the propagation and contain most of the energy in this spectral region, travel in higher-order modes, as shown in Figure~\ref{fig:Ar_sim}(b), in good agreement with our experimental results. The vertical lines in Fig.~\ref{fig:Ar_sim}(a) show the calculated phase-matching wavelengths for resonant dispersive-wave emission in the first three higher-order modes when pumping in the fundamental mode. The close agreement between the observed emission wavelength and the calculation verifies that these peaks are generated due to intermodal dispersive-wave emission.

\subsection{Supercontinuum in nitrogen-filled fiber}

We investigate the influence of Raman nonlinearity on the supercontinuum generation by filling the fiber with nitrogen instead of argon. Nitrogen exhibits similar dispersion and Kerr nonlinearity to argon; for our fiber parameters, filled with 50~bar pressure, the ratio between the nonlinear coefficients is 1.14, the soliton orders are both around 210, and the ZDWs only differ by $\sim$\SI{10}{\nm}. Figure~\ref{fig:ArvsN2}(a) shows the output spectrum for argon and nitrogen at \SI{50}{\bar} when pumping with \SI{10}{\micro\joule} pulses. The main differences between the two spectra are the spectral shape around the pump wavelength and the energy density in the normal dispersion region below \SI{835}{\nm}. The difference in the spectrum around the pump can be explained by the asymmetric self-phase modulation with enhanced red-shifting of the spectrum in nitrogen due to the Raman response~\cite{Yan1985,Nibbering1997}. However, the difference in the energy in the normal-dispersion region between argon and nitrogen requires a more detailed investigation.

Figure~\ref{fig:ArvsN2}(d) and (e) show the numerically simulated temporal evolution of the pulse in the HE$_{11}$ mode inside the fiber for argon and nitrogen respectively. In argon, the MI dynamics start after roughly \SI{25}{\cm} of propagation, with the whole pulse breaking up into a shower of few-cycle solitons~\cite{russell_hollow-core_2014}. In nitrogen, on the other hand, the second half of the pulse (i.e.~$t>0$), is strongly attenuated even before the MI dynamics start. The energy lost in the fundamental mode is coupled to higher-order modes (not shown). Since the main difference between the two cases is the presence of Raman scattering, we investigate its role more closely.


\begin{figure}[t!]
\centering
\includegraphics{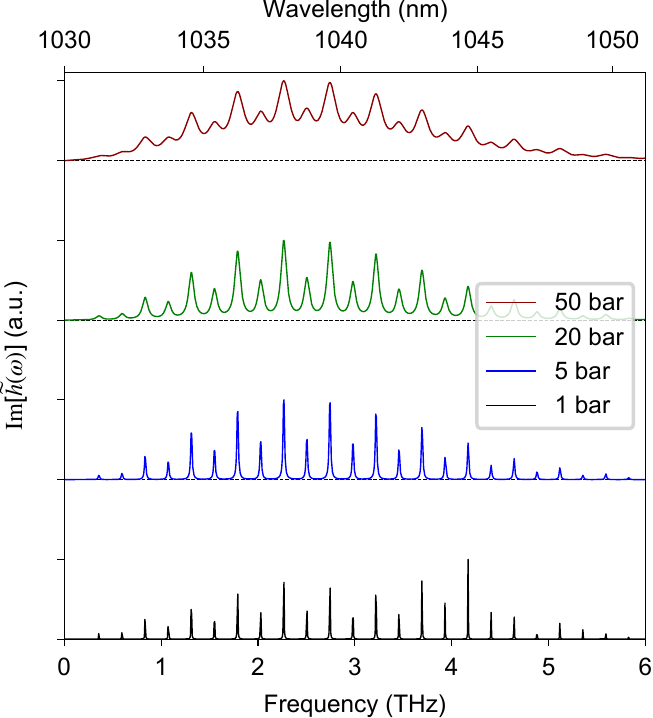}
\caption{The change in $\mathrm{Im}[\Tilde{h}(\omega)\mathrm{]}$ with pressure. Each line is normalised to its peak.}
\label{fig:gain2}
\end{figure}

\section{Discussion}

\begin{figure}[t!]
\centering
\includegraphics{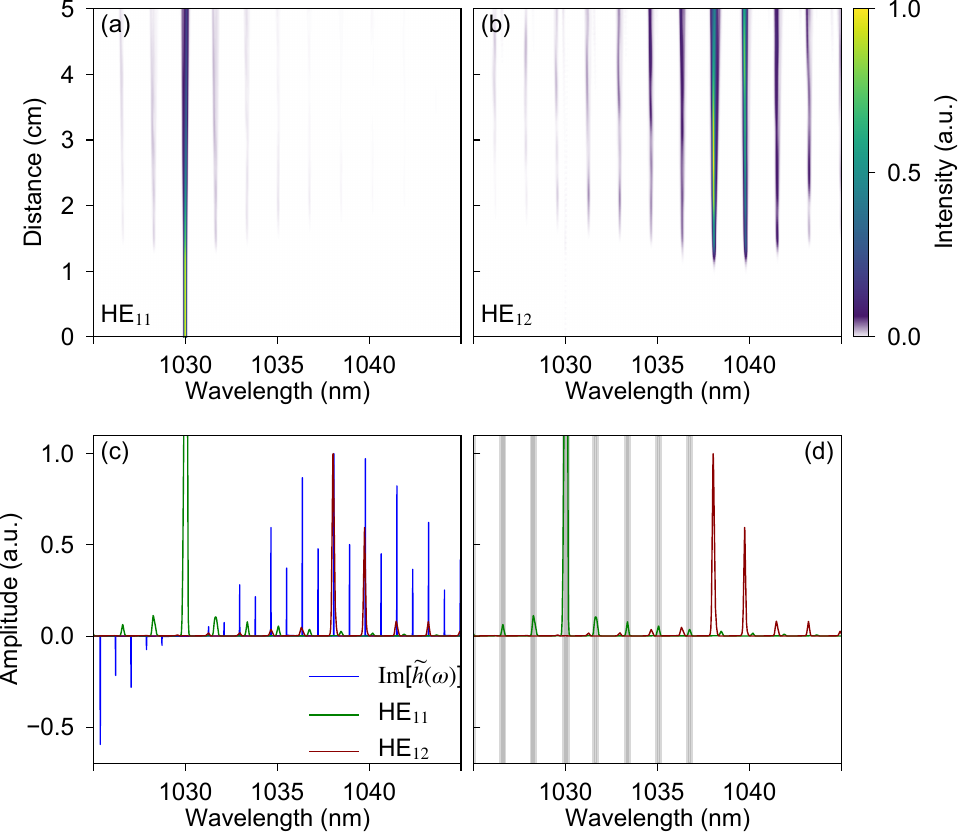}
\caption{Numerically simulated spectral evolution of a \SI{100}{\ps} transform-limited pulse propagating in \SI{50}{\bar} nitrogen with a \SI{330}{\mega\hertz} rotational linewidth. (a) and (b) Show the spectral evolution for HE$_{11}$ and HE$_{12}$ respectively. (c) Shows the spectrum of HE$_{11}$ and HE$_{12}$ at \SI{2.5}{\cm} with Stokes rotational lines response (Im[$\Tilde{h}(\omega)$]). (d) Similar to (c) but with the anti-Stokes positions marked by the grey bars, assuming they are generated by the strongest signal in HE$_{12}$.}
\label{fig:gain}
\end{figure}

Since the rotational linewidth is pressure dependent \cite{herring_temperature_1986}, at high pressure, the rotational lines broaden and merge to form a continuous Raman gain profile \cite{AGRAWAL} ($\mathrm{gain} \propto \mathrm{Im}[\Tilde{h}(\omega)]$, where $\Tilde{h}(\omega)$ is the Raman response function in the spectral domain), as shown in Figure~\ref{fig:gain2}. To simplify the simulation dynamics and to isolate the effect of the Raman response, we simulate the propagation of a narrowband pulse (\SI{100}{\ps} duration) in a fiber filled with \SI{50}{\bar} of nitrogen, but with the rotational linewidth deliberately set to \SI{330}{\mega\hertz} (which corresponds to \SI{0.1}{\bar} in the model we use \cite{gaoRamanFrequencyCombs2022}). This matches the Kerr nonlinearity and dispersion to the full model, while preserving a discrete Raman gain profile to isolate the effects of the rotational Raman response. Moreover, we switch off the vibrational response to clearly isolate the role of rotational Raman scattering. Figure~\ref{fig:gain}(a) and (b) show the simulation results for the first \SI{5}{\cm} inside the fiber. At around \SI{1}{\cm}, energy starts to be transferred from HE$_{11}$ to HE$_{12}$ through the formation of rotational Raman Stokes lines. These lines appear at the peak of the Raman gain around \SI{1038}{\nm}, as shown in Figure~\ref{fig:gain}(c). After that, more Stokes lines appear in HE$_{12}$ with their intensity proportional to the amplitude of the gain peaks. With further propagation, anti-Stokes lines generated from signals in HE$_{12}$ appear in HE$_{11}$ at and around the pump wavelength. These anti-Stokes lines coincide with the anti-Stokes line positions as shown in Figure~\ref{fig:gain}(d) (shaded bars) for the strongest signal in HE$_{12}$. After that, even more rotational lines appear through more complicated processes that involve all the rotational lines and their gain curves.

The appearance of anti-Stokes lines in a higher-order mode but not in the fundamental mode is a signature of gain suppression in the fundamental mode~\cite{loranger_sub-40_2020, hosseini_universality_2017, bloembergenCouplingVibrationsLight1964, duncan_parametric_1986}. Since the rotational shift in nitrogen is small, the propagation constants for the Stokes and anti-Stokes lines are very similar. The coherence wave created by pump-Stokes beating is thus nearly identical to the coherence wave annihilated in pump-anti-Stokes beating, causing the rate of creation and annihilation of phonons to be balanced. This results in negligible growth of either signal above the noise level \cite{bauerschmidt_dramatic_2015}. However, Stokes lines can be formed in higher-order modes by intermodal coherence waves, as observed in our simulations, because the propagation constants are sufficiently different.

We confirm the results of our simplified model experimentally by measuring the beam profile around the pump wavelength. Figure~\ref{fig:ArvsN2}(b) and (c) show the near-field mode profile after a \SI{1047}{\nm} bandpass filter for \SI{50}{\bar} argon and nitrogen respectively when pumping with \SI{2.2}{\micro\joule} of input energy. For argon, the beam profile shows a Gaussian-like fundamental mode, while for nitrogen, higher-order modes are present, as visible in the asymmetry of the beam. We do not observe the radially symmetric higher-order modes (HE$_{1m}$) as the simulation predicts. The reason for this, as we will show below, is that the coupling to higher-order modes depends on the initial coupling conditions. 

The pulses traveling in higher-order modes contribute significantly less to the formation of the supercontinuum, due to their larger anomalous dispersion and therefore lower soliton order. They can transfer energy back to the fundamental mode only while they still overlap in time with the fundamental pulse through anti-Stokes generation at the pump wavelength (using the same intermodal Stokes coherence wave). The part remaining in the higher-order mode either forms a high-order soliton and interacts nonlinearly with the gas or forms a linear wave that propagates without significant spectral broadening. In both cases, the energy transferred to the normal-dispersion region is reduced as compared to the Raman-free case.

Unlike ref.~\cite{hosseini_universality_2017}, the gain suppression here \textit{does not depend on the gas pressure}. In nitrogen, rotational lines are so closely spaced that the phase mismatch between the Stokes and anti-Stokes waves is always small. This is different from previous work in gases with widely separated rotational or vibrational Stokes lines, such as hydrogen~\cite{hosseini_universality_2017, hosseini_enhanced_2017}. We investigate this experimentally and numerically (not shown) by reducing the pressure to \SI{40}{\bar} and \SI{30}{\bar}. In both cases, we still observe higher-order-mode content in the output beam at \SI{1047}{\nm}. In addition, this mechanism of intermodal coupling when using nitrogen does not require the pulse power to be above the critical power for self-focusing, unlike in ref.~\cite{safaei_high-energy_2020}.

\begin{figure}[t!]
\centering
\includegraphics[width=1\textwidth]{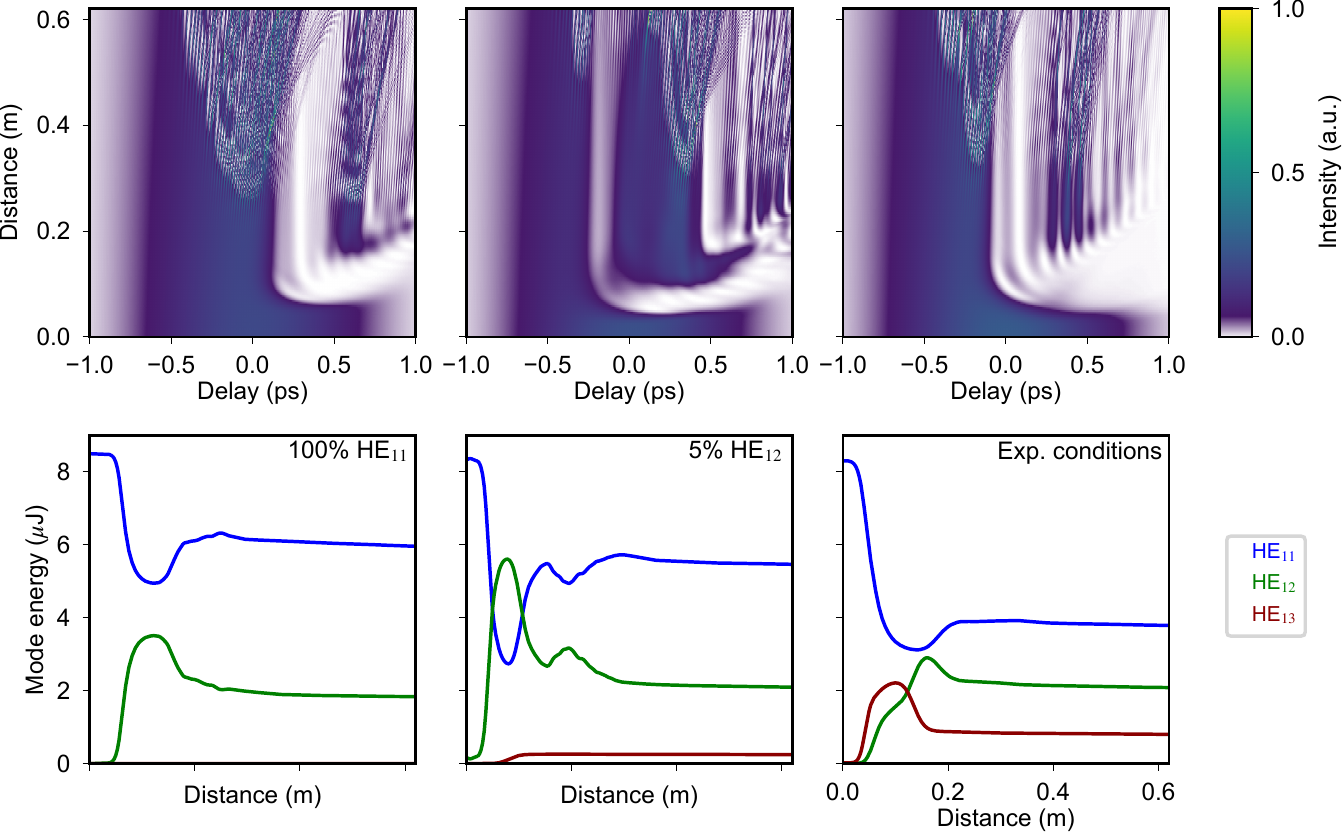}
\caption{Numerical simulation results for nitrogen with the experimental parameters in Fig.~\ref{fig:ArvsN2}(a). Top row shows the temporal evolution along the fiber for the fundamental mode HE$_{11}$. The bottom row shows the energy exchange between different HE$_{1m}$ modes with the color code shown at the top. The left column shows the case where all the input energy is coupled to HE$_{11}$. The middle column shows the case where 5\% of the input energy is coupled to HE$_{12}$. The right column shows the case where the coupling to higher-order modes is calculated based on the measured beam size at the fiber input.}
\label{fig:N2_sims}
\end{figure}

In our simulations, we find that changing the initial mode coupling changes the amount of energy exchanged between modes. Figure~\ref{fig:N2_sims} shows the temporal evolution of the pulse in HE$_{11}$ and the corresponding energy in different modes for nitrogen with the experimental parameters in Figure~\ref{fig:ArvsN2}(a) and for different input energy in the higher-order modes. In the left column, all of the input pulse energy is coupled into the fundamental mode HE$_{11}$. At \SI{12}{\cm}, around \SI{3}{\micro\joule} transfers to HE$_{12}$ through intermodal Stokes generation. With further propagation, around \SI{1}{\micro\joule} is coupled back into HE$_{11}$ through anti-Stokes generation. After that, due to the group-velocity difference between the modes, the pulses no longer overlap in time, and the energy exchange between them stops. At around \SI{20}{\cm}, the MI dynamics start with roughly \SI{6}{\micro\joule} of energy in the fundamental mode. In the middle column, we couple \SI{5}{\percent} of the input pulse energy into HE$_{12}$. This changes the amount of energy exchanged between the two modes. Moreover, while the pulses in the two modes overlap in time, some of the energy is coupled into HE$_{13}$. In the experiment, part of the input pulse is coupled into higher-order modes due to imperfect alignment and spot size at the input end of the fiber. This causes a further reduction in the supercontinuum energy density as shown in the case of the right column simulation in Figure~\ref{fig:N2_sims}.  In the right column, the coupling to higher-order modes is calculated based on the measured beam size at the fiber input face. Roughly, 98\% is coupled into the HE$_{11}$, and less than 1\% is coupled into the other modes combined. Here, even more energy is exchanged between the modes, and much less energy is transferred back to HE$_{11}$ than in the above two cases. This is because the difference in group velocity is greater between HE$_{11}$ and HE$_{13}$. Moreover, higher-order modes have higher guidance losses (not included in our model) which would cause the overall experimentally observed output energy to be lower than suggested here. Combined with the inherent energy loss due to the excitation of the gas during Raman interactions, this explains why the total output energy we observe for the nitrogen-filled case in our experiments is less than when we use argon.

\begin{figure}[t!]
\centering
\includegraphics{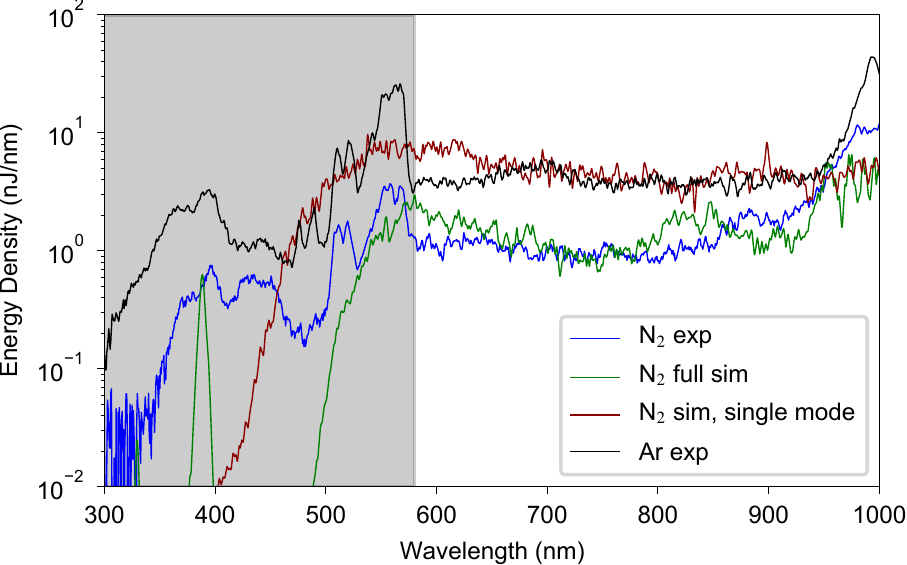}
\caption{Experimental and simulation spectrum using the experimental parameters in Fig.~\ref{fig:ArvsN2}(a). Blue: nitrogen experimental spectrum. Green: averaged spectrum of 30  full multi-mode simulations with different input noise for nitrogen  with the total input energy distributed among the modes according to the experimentally measured spot size. Red: averaged spectrum of 30 single mode simulations with different input noise for nitrogen. Black: argon experimental spectrum. The shaded indicates where the simulations and the experiment don't match due to the resonance.}
\label{fig:SimVsExp}
\end{figure}

To further validate the above results, we compare the experimental spectrum to both a full modal numerical simulation and to a single-mode simulation, to compare the impact of higher-order mode coupling. Figure~\ref{fig:SimVsExp} shows the results of these simulations along with the experimental spectrum. Above \SI{600}{\nm} (away from resonances) there is very good agreement between the full modal numerical simulation and the nitrogen experimental results. Moreover, in the single-mode simulations, in which higher-order mode coupling via gain suppression is absent, a good agreement with the argon experimental spectrum is also obtained. The discrepancy between the numerical and experimental spectrum below ~\SI{580}{\nm} is due to the dispersion model we use, which neglects the resonances caused by the fiber structure. 

\section{Summary}
In summary, we have experimentally demonstrated ultra-flat supercontinuum generation in argon-filled hollow-core fiber. We have investigated the role of the Raman response in the generation dynamics by comparing continua generated in argon and nitrogen and observed that the energy density in the normal dispersion region is lower when Raman scattering is present. We have numerically analyzed this and found that due to gain suppression of the rotational Stokes generation, a significant portion of the pulse energy is transferred from the fundamental to higher-order modes, which reduces the efficiency of supercontinuum generation. Lastly, we have found that the strength of intermodal coupling depends sensitively on the experimental conditions. 

Note that the dynamics we describe here are different from recent work on Raman supercontinuum generation~\cite{gaoRamanFrequencyCombs2022}. In the current work, the pump is located in the anomalous dispersion region, and hence MI dynamics will dominate, whereas in ref.~\cite{gaoRamanFrequencyCombs2022} the pump is in the normal dispersion which minimizes the effect of MI, and instead the supercontinuum is generated by a cascaded process involving vibrational and rotational Raman scattering as well as the Kerr effect. However, in both cases the use of nitrogen, with its closely spaced rotational Raman lines, means that gain suppression will occur. Therefore the use of alternative molecular gases, either without a rotational Raman response (such as SF$_6$ or CH$_4$) or with widely spaced lines (such as H$_2$) should be preferred.


\bibliography{ref}

\end{document}